\documentclass[journal]{IEEEtran}
\IEEEoverridecommandlockouts

\usepackage{cite}
\usepackage{amsmath,amssymb,amsfonts}
\usepackage{algorithm}
\usepackage{algpseudocode}
\usepackage{graphicx}
\usepackage{caption}
\usepackage{subcaption}
\usepackage[compact]{titlesec}  
\titlespacing{\section}{2pt}{2pt}{2pt}
\usepackage{textcomp}
\usepackage{xcolor}
\usepackage{comment}
\usepackage{hyperref}
\def\BibTeX{{\rm B\kern-.05em{\sc i\kern-.025em b}\kern-.08em
    T\kern-.1667em\lower.7ex\hbox{E}\kern-.125emX}}
\usepackage{bm}

\begin{document}

\title{Latency Optimization for Wireless Federated Learning in Multihop Networks
}

\author{\IEEEauthorblockN{Shaba Shaon, Van-Dinh Nguyen, Dinh C. Nguyen} \vspace{-15pt}
\thanks{Copyright (c) 2025 IEEE. Personal use of this material is permitted. However, permission to use this material for any other purposes must be obtained from the IEEE by sending a request to pubs-permissions@ieee.org.}
\thanks{Shaba Shaon and Dinh C Nguyen are with ECE Department,  University of Alabama in Huntsville, Huntsville, AL 35899, USA, emails: (ss0670@uah.edu, dinh.nguyen@uah.edu). Van-Dinh Nguyen is with VinUniversity, Vietnam (email: dinh.nv2@vinuni.edu.vn).}
}

\maketitle

\begin{abstract}
In this paper, we study a novel latency minimization problem in wireless federated learning (FL) across multi-hop networks. The system comprises multiple routes, each integrating leaf and relay nodes for FL model training. We explore a personalized learning and adaptive aggregation-aware FL (PAFL) framework that effectively addresses data heterogeneity across participating nodes by harmonizing individual and collective learning objectives. We formulate an optimization problem aimed at minimizing system latency through the joint optimization of leaf and relay nodes, as well as relay routing indicator. We also incorporate an additional energy harvesting scheme for the relay nodes to help with their relay tasks. This formulation presents a computationally demanding challenge, and thus we develop a simple yet efficient algorithm based on block coordinate descent and successive convex approximation (SCA) techniques. Simulation results illustrate the efficacy of our proposed joint optimization approach for leaf and relay nodes with relay routing indicator. We observe significant latency savings in the wireless multi-hop PAFL system, with reductions of up to 69.37\% compared to schemes optimizing only one node type, traditional greedy algorithm, and scheme without relay routing indicator.  
\end{abstract}
\begin{IEEEkeywords}
Federated learning, wireless, latency
\end{IEEEkeywords}

\section{Introduction}
Federated learning (FL) has appeared as an attractive solution to train machine learning (ML) models across distributed devices without data sharing \cite{12}. Despite significant milestones in FL during recent years, several fundamental challenges are yet to be addressed. In FL, model training involves frequent model exchange between servers and a large number of users. This significantly affects the FL performance as both local training and wireless transmission introduce delay. Recent efforts have been devoted to wireless FL research. In \cite{1}, the authors presented a framework with in-network aggregation to accelerate FL model training, by jointly optimizing model aggregation, routing, and spectrum allocation. The authors in \cite{2} proposed a  machine learning-enabled wireless multi-hop FL framework, while \cite{3} studied  hierarchical FL with adaptive grouping to select clients and appoint group leaders based on their ability to upload aggregated parameters to the central server. In \cite{4}, the objective is to assist the routing protocol in learning to anticipate future network topologies, and  \cite{5} investigated the impact of jamming attacks on multi-hop FL. Although there are several works that take FL as well as multi-hop networks into consideration, \textit{the latency minimization problem on wireless FL for multi-hop networks has not been investigated.}
Most of the existing works in this research area shed light on single-hop wireless networks \cite{1}. Multi-hop wireless network can provide its users with significant advantages including efficient communication, larger coverage, as well as flexibility in network reconfiguration. Overall, approaches such as energy and latency minimization can address the aforementioned problem; however, we focus on minimizing FL latency to enhance performance in wireless networks.


 
Motivated by the aforementioned challenges, \textit{this paper studies latency minimization for wireless FL over multi-hop networks}. Specifically, the contributions of this paper are three-fold:  (1) Our research explores a personalized FL framework that efficiently manages data heterogeneity among nodes by aligning individual and shared learning objectives; (2) We develop a new latency minimization problem for wireless FL over multi-hop networks by jointly considering the cooperation of leaf and  relay nodes in the FL model training. To reduce the strain on the resource-constrained relay nodes, an efficient energy harvesting scheme is integrated, enabling relay nodes to harvest energy from a portion of the radio frequency (RF) signals; (3) The latency minimization formulation results in a challenging computational problem to be solved, and thus we propose an efficient optimization solution based on  block coordinate descent (BCD) and successive convex approximation (SCA) techniques.

\section{System Model and Problem Formulation}

\subsection{Personalized FL Model}
In this work, we explore a personalized learning and adaptive aggregation-aware FL (PAFL) framework where $U$ distributed clients (nodes) train an ML model in a decentralized manner. In our multihop PAFL setup, each node $u$ trains its local model and then routes the updated local model parameters through the multihop network to the server for aggregation, as detailed in the following section. During local iteration $t$ at global round $k$, where $0 < t \leq T$ and $0 < k \leq K$, the local model training at node $u$ adheres to the following update rule:
\begin{equation} \vspace{-10pt}
    \boldsymbol{w}_{u,k}^{t+1} = \boldsymbol{w}_{u,k}^{t} - \eta \left[g_{u,k}^t + \lambda(\boldsymbol{w}_{u,k}^{t} - \boldsymbol{w}_{k})\right], \label{eqn1new}
\end{equation}
where $\boldsymbol{w}$ represents model parameters, $\eta$ denotes learning rate, $g_{n,k}^{t}$ refers to corresponding gradient, and $\lambda > 0$ is a parameter that regulates the interpolation of global and individual models. In \eqref{eqn1new}, the models are updated not only based on local gradients but also by interpolating with global parameters, effectively addressing data heterogeneity across participating nodes by harmonizing individual and collective learning objectives. Then we develop an adaptive aggregation mechanism where the model parameters from each client are weighted by its corresponding weight in the following way:
\[
\boldsymbol{w}_{k+1} = \frac{\sum_{u=1}^{U} \alpha_{u} \boldsymbol{w}_{u,k}^{T}}{\sum_{u=1}^{U} \alpha_{u}},
\]
where $\alpha_{u}$ represents the weight of each client.
\subsection{System Latency Modeling}
We consider a wireless multi-hop network where we have $U$ mobile devices (nodes) categorized into two types: leaf nodes and relay nodes. The network consists of $R$ routes originating from leaf nodes to the server. The total numbers of leaf nodes and relay nodes in the system are denoted as $M$ and $N$, respectively. We express each leaf node as $m$ and each relay node as $n$. The total number of leaf nodes in the system equals the total number of routes, i.e., $M=R$. We denote the set of all the routes in our system as $\mathcal{R}=\{1, 2, \dots ,R\}$. The sets of all the leaf nodes and all the relay nodes present in the system are expressed as $\mathcal{M}=\{1, 2, \dots ,M\}$ and $\mathcal{N}=\{1, 2, \dots ,N\}$, respectively. Each route $r$ may accommodate a different number of relay nodes during global round $k$, i.e., each route comprises one leaf node and several relay nodes. We introduce a variable to model the uncertainty due to nodes' mobility and route availability, and this is commonly done using a binary routing indicator. During global round $k$, we express the routing indicator for relay node $n$ as $\delta_{n}^{r,k}$. More specifically,
\[
  \delta_{n}^{r,k} =
  \begin{cases}
    1, & 
      \begin{aligned}[t]
        &\text{if a valid route $r$ exists for relay node } n \text{ in } \\
        &\text{ global round } k,
      \end{aligned}\\[6pt]
    0, & 
      \begin{aligned}[t]
        &\text{if the node belongs to any other route or} \\
        &\text{is in routing outage in round } k. 
      \end{aligned}
  \end{cases}
\]

This indicates whether relay node $n$ is connected to the server in global round $k$.
In a wireless `ad-hoc' network, each node participates in routing by forwarding data to other nodes. In our model, leaf nodes train and upload their local models to their immediate relay node. Relay nodes train, upload their models, and relay local models of all the nodes they are predecessors to. 

For leaf node $m$ during global round $k$, let $f_{m}^{k}$ represent its CPU computation capability (in CPU cycles per second), $D_{m}^{k}$ denote the number of data samples, and $C_{m}^{k}$ stand for the number of CPU cycles needed to process a data sample. If $L_{m}^{k}$ denotes the number of local iterations, the computation time during global round $k$ is calculated as $T_{m}^{\text{train},k}=\frac{L_{m}^{k} C_{m}^{k} D_{m}^{k}}{f_{m}^{k}}$. The corresponding energy consumption is given by $E_{m}^{\text{train},k}=L_{m}^{k} \zeta_{m} C_{m}^{k} D_{m}^{k} {{f_{m}^{k}}^{2}}$, where $\zeta_{m}$ depends on the hardware and chip architecture of leaf node $m$ \cite{12}. After local computation, each user uploads its updated local model parameters to the server for aggregation. We employ frequency division multiple access for the uplink operation. For leaf node $m$, the achievable rate during global round $k$ is determined by $R_{m}^{k} =b_{m}^{k}\log_{2}\left(1+\frac{p_{m}^{k}g_{m}^{k}}{b_{m}^{k}n_{0}}\right)$, where $b_{m}^{k}$ represents the allocated bandwidth, $p_{m}^{k}$ is the transmit power, $g_{m}^{k}$ stands for the channel gain of leaf node $m$, and $n_{0}$ denotes the noise power spectral density. Assuming a constant data size $s$ for the local model parameters, the uploading time can be expressed as, as $T_{m}^{\text{up},k}=\frac{s}{R_{m}^{k}}$, and the corresponding energy consumption is $E_{m}^{\text{up},k}=T_{m}^{\text{up},k}p_{m}^{k}$. Hence, the total time $T_{m}$ required for computing and uploading local model parameters for leaf node $m$ during global round $k$ is $T_{m}^{k}=T_{m}^{\text{train},k}+T_{m}^{\text{up},k}$. If the total energy consumed by leaf node $m$ for computing and uploading local models during each global iteration is denoted by $E_{m}^{k}$, it can be expressed as $E_{m}^{k}=E_{m}^{\text{train},k}+E_{m}^{\text{up},k}$.

For relay node $n$ during global round $k$, the computation time for $L_{n}^{k}$ local iterations is calculated as $T_{n}^{\text{train},k}=\frac{L_{n}^{k}C_{n}^{k}D_{n}^{k}}{f_{n}^{k}}$. Here, $f_{n}^{k}$ represents the CPU computation capability (in CPU cycles per second), $D_{n}^{k}$ denotes the number of data samples, and $C_{n}^{k}$ stands for the number of CPU cycles needed to process a data sample. The corresponding energy consumption by relay node $n$ is given by $E_{n}^{\text{train},k}=L_{n}^{k}\zeta_{n}C_{n}^{k}D_{n}^{k}{{f_{n}^{k}}^{2}}$, where $\zeta_{n}$ depends on the hardware and chip architecture of relay node $n$ \cite{12}. Moreover, $\delta_{n}^{k}$ is the binary routing indicator for relay node $n$ that specifies whether the node is connected to the server through any route in round $k$. Similar to leaf nodes, after local computation, relay nodes upload their local models to the server for aggregation. The uploading time for relay node $n$ during global round $k$ is given by $T_{n}^{\text{up},k}=\frac{\delta_{n}^{r,k}s}{R_{n}^{k}}$, where $s$ represents the constant data size of the local model parameters uploaded by relay node $n$. The achievable uploading rate $R_{n}^{k}$ is determined by $R_{n}^{k}=b_{n}^{k}\log_{2}\left(1+\frac{p_{n}^{k}g_{n}^{k}}{b_{n}^{k}n_{0}}\right)$, where $b_{n}^{k}$ stands for the allocated bandwidth, $p_{n}^{k}$ denotes the transmit power, and $g_{n}^{k}$ represents the channel gain of relay node $n$ during global round $k$. The corresponding energy consumption is expressed as $E_{n}^{\text{up},k}=T_{n}^{\text{up},k}p_{n}^{k}$. In this work, all channels are assumed to have two types of fading effects that characterize mobile wireless communications: large-scale fading and small-scale fading. The small-scale fading component is modeled using a Rayleigh distribution, while the large-scale fading coefficient is represented by a deterministic path loss model which is discussed later in the Energy Harvesting Scheme section.

Additionally, a relay node must transmit the local models of all nodes it precedes. We assume that within a route, a relay node $n$ is connected to several successor nodes, i.e. one leaf node and $n'$ relay nodes. Let $T_{n}^{\text{tx},k}$ represent the time required by relay node $n$ for transmitting all the local models of $n'$ relay nodes it precedes, where $T_{n}^{\text{tx},k}=\sum_{i=1}^{n'} T_{n,i}^{\text{tx},k}$. Similarly, if $T_{n,m}^{\text{tx},k}$ stands for the time required for transmitting the local model of one leaf node it precedes, then $T_{n,m}^{\text{tx},k}=\frac{s}{R_{n}^{k}}$. The energy consumption by relay node $n$ to transmit the local models of all the nodes it precedes is $E_{n}^{\text{tx},k}=E_{n}^{\text{up},k}+(n')E_{n}^{\text{up},k}=(1+n')E_{n}^{\text{up},k}$. This equation yields from our assumption of same local model size for all the nodes. Because of this assumption, energy consumption for uploading local model parameters of size $s$ depends on the achievable uploading rate and transmit power of the acting node. That is why, for relay node $n$, it takes the same amount of energy to transmit the local model parameters of each of the nodes it precedes. Thus, the time $T_{n}^{k}$ required by relay node $n$ to compute, upload and transmit during global round $k$ is expressed as $T_{n}^{k}=T_{n}^{\text{train},k}+T_{n}^{\text{up},k}+T_{n,m}^{\text{tx},k}+T_{n}^{\text{tx},k}$. Similarly, the corresponding energy consumption by relay node $n$ to compute, upload and transmit can be written as $E_{n}^{k}=E_{n}^{\text{train},k}+E_{n}^{\text{up},k}+E_{n}^{\text{tx},k}$. 

If $T_{\text{total}}^{r}$ is the total time required for route $r$ to complete global round $k$, then it is formulated as $T_{\text{total}}^{r,k}=(T_{m}+\sum_{n=1}^{N} T_{n}^{k})$. As the route that takes the longest time to complete each global iteration will be the bottleneck for the latency, the total time required for completing global round $k \in \mathcal{K}=\{1, 2, \dots, K\}$ is written as $T_{\text{total}}^{k} = \underset{r \in \mathcal{R}}{\max} T_{\text{total}}^{r,k}  = \underset{r \in \mathcal{R}}{\max} (T_{m}^{k}+\sum_{n=1}^{N} T_{n}^{k})$. Hence, the total latency of the FL system over $K$ global rounds can be expressed as
\begin{equation} \vspace{-10pt}
T_{\text{total}}^{\text{FL}} = \sum_{k=1}^{K}\left(T_{\text{total}}^{k} \right)  = \sum_{k=1}^{K}\left(\underset{r \in \mathcal{R}}{\max} (T_{m}^{k}+\sum_{n=1}^{N} T_{n}^{k})\right).
\end{equation}

To further support sustainable FL, we propose energy harvesting (EH) inspired by \cite{6}, where the received RF signal at relay node $n$ from the previous node is given as $y_{n}=\sqrt{p_{n-1}}g_{n}\hat{x}_{n}+N_{n}, n=1, 2, ... , N+1$, where $p_{n-1}$ is the transmit power of relay node $(n-1)$, $g_{n}$ is the channel gain between current and previous relay node, and $\hat{x}_{k}$ is the information signal from the previous relay node. The channel gain $g_{n}$ can be modeled as $g_{n}=\sqrt{\xi_{n}}\Tilde{g}_{n}$, where, $\xi_{n}$ is the large-scale fading coefficient, $\Tilde{g}_{n}$ represents the small-scale fading component with Rayleigh distribution. The large scale fading coefficient can be modeled as $\xi_{n}=A_{n}(\frac{d_{n}}{d_{0}})^{-\alpha_{n}}$, where $A_{n}$ is the reference attenuation at a reference distance of $d_{0}$. $d_{n}$ represents the distance between transmit and receive relay nodes. $\alpha_{n}$ is the path loss exponent. Then, this received RF signal at relay node $n$ is split into
two for harvesting energy (EH) as well as decoding and transmitting information (ID) based on the PS ratio, $\rho_{k}$. The EH signal at relay node $n$ can be written as $y_{n}^{EH}=\sqrt{\rho_{k}}\left(\sqrt{p_{n-1}}g_{n}\hat{x}_{n}+N_{n}\right)$. Similarly, the ID signal at relay node $n$ can be written as $y_{n}^{EH}=\sqrt{(1-\rho_{k})}\left(\sqrt{p_{n-1}}g_{n}\hat{x}_{n}+N_{n}\right)+z_{n}$, where, $z_{k}$ represents the additional noise introduced by ID circuitry.
Thus, the harvested energy at relay node $n$ can be expressed as $E_{n}^{EH}=\beta_{n} \underset{\hat{x}_{n},N_{n}}{\mathbb{E}}[|y_{n^{EH}}|^{2}]\approx\beta_{n}\rho_{n}E_{n-1}|g_{n}|^{2} = E_{0}\lambda_{n}\overset{n}{\underset{j=1}{\prod}}\rho_{j}, n=1, 2, ... , N+1$, where $\lambda_{n}=\overset{n}{\underset{j=1}{\prod}}\beta_{j}|g_{j}|^{2}$ and $0 < \beta_{n} \le 1$ is the energy conversion efficiency of relay node $n$. Now, if relay node $n$ has its own energy resource $E_{n}^{self}$ for its own computation and communication, then the total usable energy of relay node $n$ can be expressed as $E_{n}^{\max}=E_{n}^{\text{self}}+E_{n}^{\text{EH}}$.
 
\subsection{Problem Formulation}
This research aims to minimize the latency of the FL algorithm. Based on the above analysis, we formulate the following optimization problem:
\begin{subequations} 
\begin{align}
\min_{\pmb{p_{m}^{k}}, \pmb{f_{m}^{k}}, \pmb{p_{n}^{k}}, \pmb{f_{n}^{k}}, \pmb{\delta_{n}^{k,r}}} \quad & T_{\text{total}}^{\text{FL}} \label{eqn:2a}\\ 
\textrm{s.t.} \quad & 0 \le p_{m}^{k} \le P_{m}, \forall{m} \label{eqn:2b}\\
& 0 \le p_{n}^{k} \le P_{n}, \forall{n}  \label{eqn:2c}\\
& 0 \le f_{m}^{k} \le F_{m}, \forall{m}  \label{eqn:2d}\\
& 0 \le f_{n}^{k} \le F_{n}, \forall{n} \label{eqn:2e}\\
& E_{m}^{k} \le E_{m}^{\max}, \forall{m} \label{eqn:2f}\\
& E_{n}^{k} \le E_{n}^{\max}, \forall{n} \label{eqn:2g}\\
& \delta_{n}^{r,k} \in \{0,1\}, \forall{n}, \forall{r}, \forall{k}. \label{eqn:2h}
\end{align}
\end{subequations}
\noindent
where $\pmb{p_{m}}=\{p_{1}, p_{2}, \dots ,p_{M}\}$, $\pmb{p_{n}}=\{p_{1}, p_{2}, \dots ,p_{N}\}$, $\pmb{f_{m}}=\{f_{1}, f_{2}, \dots ,f_{M}\}$, $\pmb{f_{n}}=\{f_{1}, f_{2}, \dots ,f_{N}\}$, and $\pmb{\delta_{n}^{k,r}}=\{\delta_{1}^{1,1}, \delta_{2}^{1,1}, \dots ,\delta_{N}^{K,R}\}$. In (2), \eqref{eqn:2b} and \eqref{eqn:2c} represent the feasible range of the transmit power due to the power budgets of the leaf nodes and the relay nodes. The CPU frequency of each node is constrained in \eqref{eqn:2d} and \eqref{eqn:2e}. The constraints \eqref{eqn:2f} and \eqref{eqn:2g} are on the energy consumption by each leaf node and relay node, respectively. \eqref{eqn:2h} is on the binary routing indicator for relay nodes. 

\section{Proposed Solution}
Solving problem in (3) directly is a challenging task as multiple optimization variables are coupled. The objective function \eqref{eqn:2a} as well as the energy constraints \eqref{eqn:2f} and \eqref{eqn:2g} are non-convex in nature because of the presence of $\log_{2}$ function of the achievable rates. Moreover, the binary routing indicator constraint \eqref{eqn:2h} is not continuous. To overcome the non-convex nature of the objective function and the aforementioned constraints, we divide problem in (2) into three sub-problems. Hence, the control variables of problem in (2) are divided into three blocks: (i) the first block is for binary routing indicator optimization $(\delta_{n}^{k,r})$ for relay nodes, (ii) the second block is for leaf node optimization $(p_{m}, f_{m})$ and (iii) the third block for relay node optimization $(p_{n}, f_{n})$, which will be updated alternatively in an iterative manner.

\textit{\textbf{For the first block}}, \textbf{problem in (2)} is equivalently re-written as \vspace{-15pt}
\begin{subequations} 
\begin{align}
\min_{\delta_{n}^{k,r}} \quad & \sum_{k=1}^{K}\left[\max_{r \in \mathcal{R}} \left({\frac{L_{m}^{k}C_{m}^{k}D_{m}^{k}}{f_{m}^{k}}}+\frac{s}{b_{m}^{k}\log_{2}\left(1+\frac{p_{m}^{k}g_{m}^{k}}{b_{m}^{k}n_{0}}\right)}\right.\right. \nonumber\\
& \left.\left.+\sum_{n=1}^{N}\left({\frac{L_{n}C_{n}D_{n}}{f_{n}}}+\frac{\delta_{n}^{r,k}(n'+2)s}{b_{n}^{k}\log_{2}\left(1+\frac{p_{n}^{k}g_{n}^{k}}{b_{n}^{k}n_{0}}\right)}\right)\right)\right] \label{eqn:4a'}\\ 
\textrm{s.t.} \quad & 0 \leq \delta_{n}^{r,k} \leq 1, \forall{n}, \forall{r}, \forall{k}. \label{eqn:4b'}
\end{align}
\end{subequations} \vspace{-15pt}

In \eqref{eqn:4b'}, we have transformed binary routing variable of relay nodes into a continuous variable. Since the problem in (4) is already convex, it can be solved directly using convex optimization problem solvers. \textit{For complexity analysis}, this problem consists of $(N)$ scalar decision variables and $(N)$ linear constraints, which results in the per-iteration computational complexity of $\mathcal{O}\left((N)^2\sqrt{N}\right)$ \cite{ben2001lectures}.

\textit{\textbf{For the second block}},  let us introduce a new slack variable $x_{m}^{k}$ such that:

\vspace{-15pt}
\begin{equation}
x_{m}^{k} \geq \frac{s}{b_{m}^{k}\log_{2}\left(1+\frac{p_{m}^{k}g_{m}^{k}}{b_{m}^{k}n_{0}}\right)}, \forall{m}. 
\end{equation} 
\textbf{problem in (3)} is equivalently re-written as
\begin{subequations} 
\begin{align}
\min_{p_{m}^{k},f_{m}^{k}} \quad & \sum_{k=1}^{K}\left[\max_{r \in \mathcal{R}} \left({\frac{L_{m}^{k}C_{m}^{k}D_{m}^{k}}{f_{m}^{k}}}+x_{m}^{k}\right.\right. \nonumber\\
& \left.\left.+\sum_{n=1}^{N}\left({\frac{L_{n}^{k}C_{n}^{k}D_{n}^{k}}{f_{n}^{k}}}+\frac{\delta_{n}^{r,k}(n'+2)s}{b_{n}^{k}\log_{2}\left(1+\frac{p_{n}^{k}g_{n}^{k}}{b_{n}^{k}n_{0}}\right)}\right)\right)\right] \label{eqn:4a}\\ 
\textrm{s.t.} \quad & L_{m}^{k}\zeta_{m}C_{m}^{k}D_{m}^{k}{{f_{m}^{k}}^{2}}+x_{m}^{k}p_{m}^{k} \le E_{m}^{\max}, \forall{m} \label{eqn:4b}\\
& \frac{s}{b_{m}^{k}x_{m}^{k}} \leq \log_{2}\left(1+\frac{p_{m}^{k} g_{m}^{k}}{b_{m}^{k} n_{0}}\right), \forall{m} \label{eqn:4c}\\
& \eqref{eqn:2b}, \eqref{eqn:2d}. \label{eqn:4d}
\end{align}
\end{subequations}

We see that objective \eqref{eqn:4a} is convex, while constraints in \eqref{eqn:4d} are also convex. Now we focus on converting constraints \eqref{eqn:4b} and \eqref{eqn:4c} into convex ones.

\textit{Constraint \eqref{eqn:4b}}: For $x_{m}^{k}>0$ and $p_{m}^{k}>0$, we apply SCA to approximate $x_{m}^{k}p_{m}^{k}$ as

\begin{equation} \label{eqn:5}
x_{m}^{k} p_{m}^{k} \le \frac{1}{2} \frac{p_{m}^{k,i}}{x_{m}^{k,i}} {x_{m}^{k}}^{2} + \frac{1}{2} \frac{x_{m}^{k,i}}{p_{m}^{k,i}} {p_{m}^{k}}^{2} = h_{m}^{k,i}(x_{m}^{k},p_{m}^{k})
\end{equation}

\noindent
where $p_{m}^{k,i}$ and $x_{m}^{k,i}$ are the feasible point of $p_{m}^{k}$ and $x_{m}^{k}$ at iteration $i$. Hence constraint \eqref{eqn:4b} can be convexified as

\begin{equation} \label{eqn:6}
    L_{m}^{k}\zeta_{m}C_{m}^{k}D_{m}^{k}{{f_{m}^{k}}^{2}} +\frac{1}{2} \frac{p_{m}^{k,i}}{x_{m}^{k,i}} {x_{m}^{k}}^{2} + \frac{1}{2} \frac{x_{m}^{k,i}}{p_{m}^{k,i}} {p_{m}^{k}}^{2} \le E_{m}^{\max}, \forall{m}.
\end{equation}

\textit{Constraint \eqref{eqn:4c}}: We use this inequality
\begin{align} \label{eqn:34}
\ln(1+z) \ge \ln(1+z_{i}) + \frac{z_{i}}{z_{i}+1} - \frac{(z_{i})^{2}}{z_{i}+1} \frac{1}{z}.
\end{align}
Now we approximate RHS of \eqref{eqn:4c} as

\begin{equation} \label{eqn:8}
\begin{split}
    \frac{ s \ln 2}{b_{m}^{k}x_{m}^{k}} \leq \ln \left(1+\frac{p_{m}^{k,i} g_{m}^{k}}{b_{m}^{k} n_{0}}\right) + \frac{p_{m}^{k,i} g_{m}^{k}}{p_{m}^{k} g_{m}^{k}+b_{m}^{k} n_{0}} \\
    - \frac{(p_{m}^{k,i} g_{m}^{k})^{2}}{p_{m}^{k,i} g_{m}^{k} + b_{m}^{k} n_{0}} \frac{1}{p_{m}^{k} g_{m}^{k}}, \forall{m}.
\end{split}
\end{equation}

So, we solve the following convex problem at iteration $i+1$:
\begin{subequations} 
\begin{align}
\min_{p_{m}^{k},f_{m}^{k}} \quad & \sum_{k=1}^{K}\left[\max_{r \in \mathcal{R}} \left({\frac{L_{m}^{k}C_{m}^{k}D_{m}^{k}}{f_{m}^{k}}}+x_{m}^{k}\right.\right. \nonumber\\
& \left.\left.+\sum_{n=1}^{N}\left(\delta_{n}^{r,k}{\frac{L_{n}^{k}C_{n}^{k}D_{n}^{k}}{f_{n}^{k}}}+\frac{(n'+2)s}{\delta_{n}^{r,k}b_{n}^{k}\log_{2}\left(1+\frac{p_{n}^{k}g_{n}^{k}}{b_{n}^{k}n_{0}}\right)}\right)\right)\right] \label{eqn:9a}\\ 
\textrm{s.t.} \quad & \eqref{eqn:4d}, \eqref{eqn:6}, \eqref{eqn:8}. \label{eqn:9b}
\end{align}
\end{subequations}
\textit{For complexity analysis}, this problem consists of $(2M)$ scalar decision variables and $(4M)$ linear or quadratic constraints, which results in the per-iteration computational complexity of $\mathcal{O}\left((2M)^2\sqrt{4M}\right)$ \cite{ben2001lectures}.

\textit{\textbf{For the third block}}, let us introduce a new slack variable $y_{n}^{k}$ such that:
\begin{equation}
y_{n}^{k} \ge \frac{\delta_{n}^{r,k}(n'+2) s}{b_{n}^{k}\log_{2}\left(1+\frac{p_{n}^{k}g_{n}^{k}}{b_{n}^{k}n_{0}}\right)}, \forall{n}. 
\end{equation}
\textbf{problem in (3)} is equivalently re-written as

\begin{subequations} 
\begin{align}
\min_{p_{n}^{k},f_{n}^{k}} \quad & \sum_{k=1}^{K}\left[\max_{r \in \mathcal{R}} \left(\left({\frac{L_{m}^{k}C_{m}^{k}D_{m}^{k}}{f_{m}^{k}}}+\frac{s}{b_{m}^{k}\log_{2}\left(1+\frac{p_{m}^{k}g_{m}^{k}}{b_{m}^{k}n_{0}}\right)}\right)\right.\right. \nonumber\\
& \left.\left.+\sum_{n=1}^{N}\left({\frac{L_{n}^{k}C_{n}^{k}D_{n}^{k}}{f_{n}^{k}}}+y_{n}^{k}\right)\right)\right] \label{eqn:11a}\\ 
\textrm{s.t.} \quad & L_{n}^{k}\zeta_{n}C_{n}^{k}D_{n}^{k}{{f_{n}^{k}}^{2}}+y_{n}^{k}p_{n}^{k} \le E_{n}^{\max}, \forall{n} \label{eqn:11b}\\
& \frac{\delta_{n}^{r,k}(n'+2)s}{b_{n}^{k}y_{n}^{k}} \leq \log_{2}\left(1+\frac{p_{n}^{k} g_{n}^{k}}{b_{n}^{k} n_{0}}\right), \forall{n}. \label{eqn:11c}\\
& \eqref{eqn:2c}, \eqref{eqn:2e}. \label{eqn:11d} 
\end{align}
\end{subequations}

\noindent Here, the objective function \eqref{eqn:11a} and constraint in \eqref{eqn:11d} are convex. However, constraints \eqref{eqn:11b} and \eqref{eqn:11c} are still non-convex. For convexifying these two constraints, we follow the same strategy as for constraints \eqref{eqn:4b} and \eqref{eqn:4c}.

\textit{Constraint \eqref{eqn:11b}}: Similar to constraint \eqref{eqn:4b}, constraint \eqref{eqn:11b} can be convexified as
\begin{equation} \label{eqn:12}
    L_{n}^{k}\zeta_{n}C_{n}^{k}D_{n}^{k}{{f_{n}^{k}}^{2}} +\frac{1}{2} \frac{p_{n}^{k,i}}{y_{n}^{k,i}} {y_{n}^{k}}^{2} + \frac{1}{2} \frac{y_{n}^{k,i}}{p_{n}^{k,i}} {p_{n}^{k}}^{2} \le E_{n}^{max}, \forall{n}
\end{equation}
\noindent
where $p_{n}^{k,i}$ and $y_{n}^{k,i}$ are the feasible point of $p_{n}^{k}$ and $y_{n}^{k}$ at SCA iteration $i$. 

\textit{Constraint \eqref{eqn:11c}}: Similar to  constraint \eqref{eqn:4c}, we approximate RHS of \eqref{eqn:11c} as
\begin{equation} \label{eqn:13}
\begin{split}
    \frac{\delta_{n}^{r,k}(n'+2) s \ln{2}}{ b_{n}^{k} y_{n}^{k}} \le \ln \left(1+\frac{p_{n}^{k,i} g_{n}^{k}}{b_{n}^{k} n_{0}}\right) + \frac{p_{n}^{k,i} g_{n}^{k}}{p_{n}^{k} g_{n}^{k}+b_{n}^{k} n_{0}} \\
    - \frac{(p_{n}^{k,i} g_{n}^{k})^{2}}{p_{n}^{k,i} g_{n}^{k} + b_{n}^{k} n_{0}} \frac{1}{p_{n}^{k} g_{n}^{k}}, \forall{n}.
\end{split}
\end{equation}
Thus, we solve the following convex problem at iteration $i+1$:
\begin{subequations} 
\begin{align}
\min_{p_{n}^{k},f_{n}^{k}} \quad & \sum_{k=1}^{K}\left[\max_{r \in \mathcal{R}} \left(\left({\frac{L_{m}^{k}C_{m}^{k}D_{m}^{k}}{f_{m}^{k}}}+\frac{s}{b_{m}^{k}\log_{2}\left(1+\frac{p_{m}^{k}g_{m}^{k}}{b_{m}^{k}n_{0}}\right)}\right)\right.\right. \nonumber\\
& \left.\left.+\sum_{n=1}^{N}\left({\frac{L_{n}^{k}C_{n}^{k}D_{n}^{k}}{f_{n}^{k}}}+y_{n}^{k}\right)\right)\right] \label{eqn:14a}\\ 
\textrm{s.t.} \quad & \eqref{eqn:11d}, \eqref{eqn:12}, \eqref{eqn:13}. \label{eqn:14b}
\end{align}
\end{subequations}
\textit{For complexity analysis}, this problem consists of $(2N)$ scalar decision variables and $(4N)$ linear or quadratic constraints, which results in the per-iteration computational complexity of $\mathcal{O}\left((2N)^2\sqrt{4N}\right)$ \cite{ben2001lectures}.
To summarize, we jointly solve the above three blocks to obtain the solutions for \textbf{problem in (3)}, as illustrated in Algorithm 1.
\makeatletter
\renewcommand{\Statex}{\item[]\hskip\ALG@thistlm}
\makeatother

\begin{algorithm}[t]
  \caption{SCA-based Optimization Algorithm}
\begin{algorithmic}[1]
 \footnotesize
    \Statex \textbf{Input:} 
            \Statex Set the iteration index $i=0$;
            \Statex Initialize a feasible solution
            (${\delta_{n}^{r,k}}^{0}$, ${p_{m}^{k}}^{0}$, ${f_{m}^{k}}^{0}$, ${p_{n}^{k}}^{0}$, ${f_{n}^{k}}^{0}$) for the problem in (3);
    \Statex \textbf{Repeat}
            \Statex Set $i \gets i+1$
            \Statex Solve problem in (4) to update $\delta_{n}^{r,k}$;
            \Statex Solve problem in (11) to update $p_{m}^{k,i}$, $f_{m}^{k,i}$;
            \Statex Solve problem in (16) to update $p_{n}^{k,i}$, $f_{n}^{k,i}$;
    \Statex \textbf{Until} convergence. 
    \Statex \textbf{Output:}
            \Statex Optimal $\pmb{{\delta_{n}^{r,k}}^{*}}$ ,$\pmb{{p_{m}^{k}}^{*}}$, $\pmb{{f_{m}^{k}}^{*}}$, $\pmb{{p_{n}^{k}}^{*}}$, $\pmb{{f_{n}^{k}}^{*}}$.
\end{algorithmic}
\end{algorithm}
\vspace{-3pt}
\section{Simulation Results and Evaluation}
A multi-hop wireless communication environment has been considered that consists of three routes. Route 1, 2, and 3 each consist of a varying number of relay nodes, which are assigned based on the relay routing indicator during each global round, with one leaf node assigned to each route. We have considered practical scenarios for simulation \cite{12}. The system bandwidth is considered to be 20 MHz \cite{12}. The maximum transmit power $P_{m}$ of leaf nodes and $P_{n}$ of relay nodes are configured in the range of [5-25] dBm. The noise power density is set to $N_{0}$= -174 dBm/Hz \cite{12}. The maximum CPU cycle frequency of a leaf node is configured as $F_{m}$ = 2GHz and that of a relay node is also configured as $F_{n}$ = 2GHz \cite{12}. The coefficients for leaf and relay nodes, which are contingent on their respective hardware and chip architecture, are established as $\zeta_{m} = 10^{-28}$ and $\zeta_{n} = 10^{-28}$, respectively \cite{12}. The number of local iterations for leaf nodes is considered to be $L_{m}$=5, while that for relay nodes is considered to be $L_{n}$ = 15. All simulations were conducted in Matlab using YALMIP toolbox with the solver MOSEK. To demonstrate the effectiveness of our joint leaf-relay node with relay routing indicator optimization method, we compare our proposed scheme with four baselines: (i) Scheme 1-optimization for only leaf nodes, (ii) Scheme 2-optimization for only relay nodes, (iii)Scheme 3-optimization for both leaf and relay nodes without relay routing indicator and (iii) Greedy.


\begin{figure}[ht!]
    \centering
    \footnotesize
    \begin{subfigure}[t]{0.49\linewidth} 
        \centering
        \includegraphics[width=\linewidth]{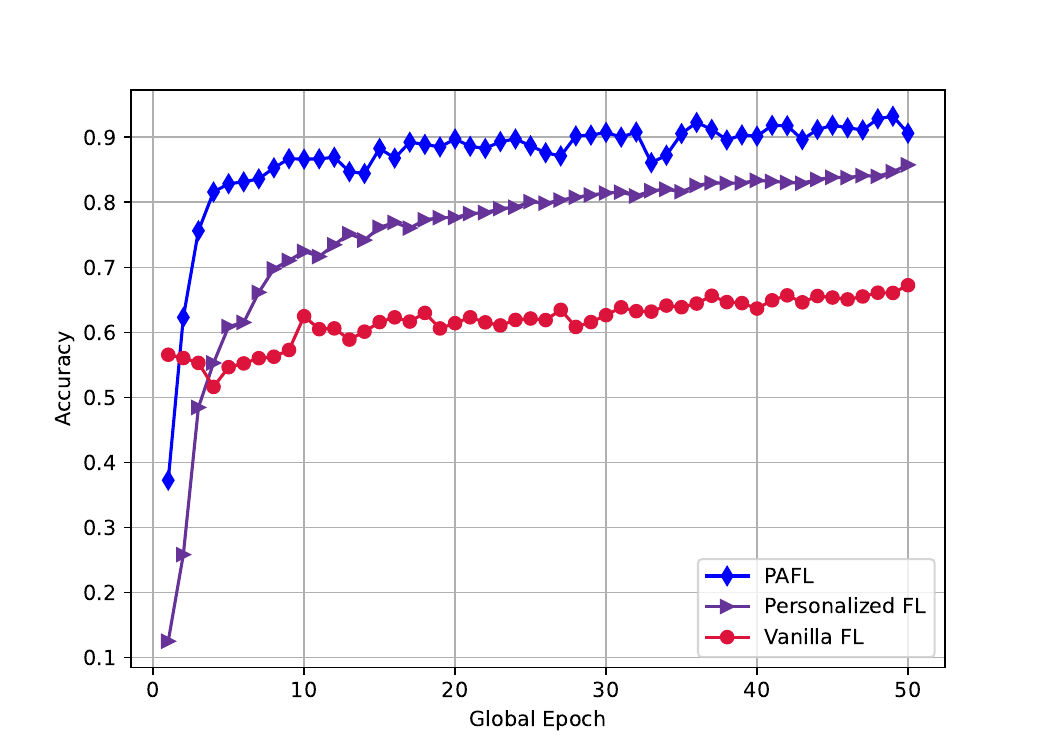}
        \caption{Average accuracy versus global communication rounds of FL model training for MNIST dataset (non-IID).}
        \label{fig:sub1}
    \end{subfigure}
    \hfill 
    \begin{subfigure}[t]{0.49\linewidth} 
        \centering
        \includegraphics[width=\linewidth]{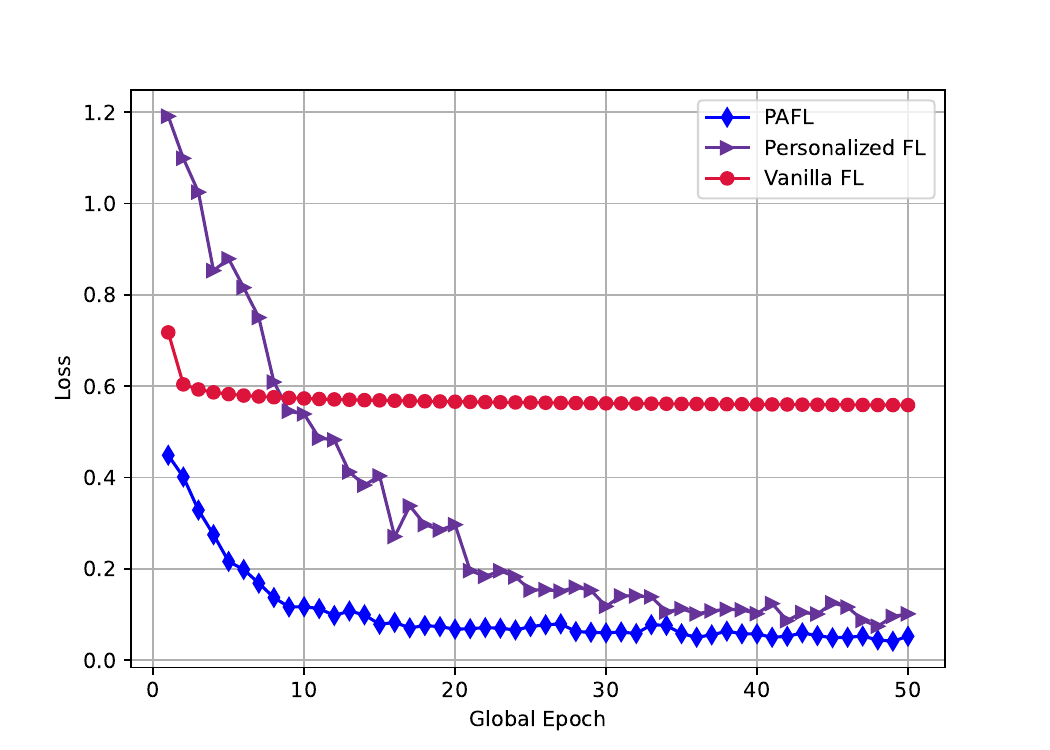}
        \caption{Average loss versus global communication rounds of FL model training for MNIST dataset (non-IID).}
        \label{fig:sub2}
    \end{subfigure}
        \begin{subfigure}[t]{0.49\linewidth} 
        \centering
        \includegraphics[width=\linewidth]{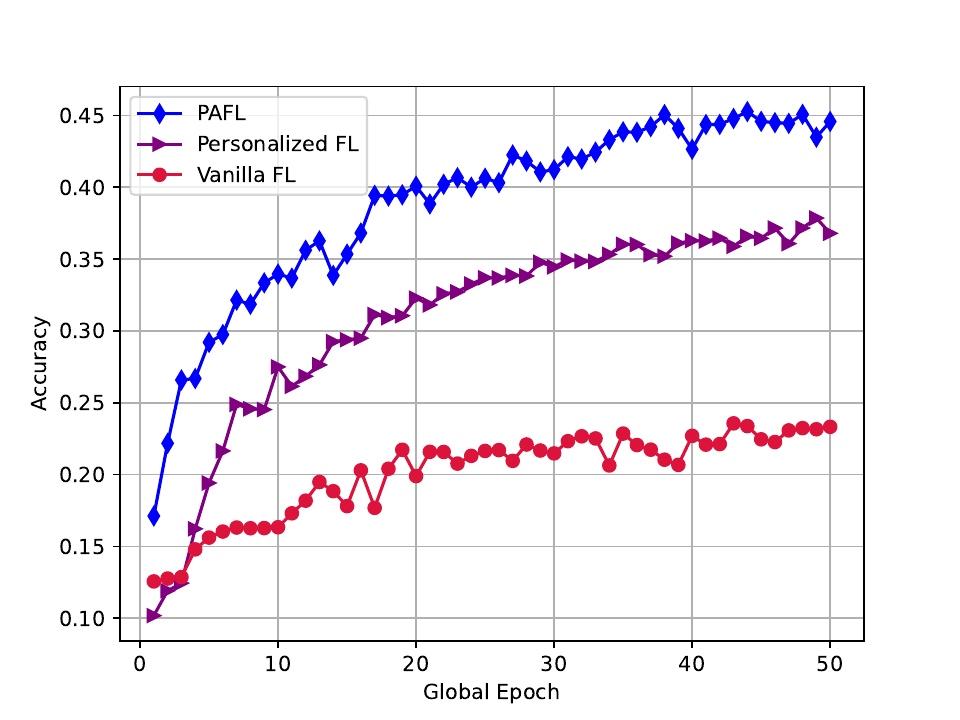}
        \caption{Average accuracy versus global communication rounds of FL model training for Cifar-10 dataset (non-IID).}
        \label{fig:sub3}
    \end{subfigure}
    \hfill 
    \begin{subfigure}[t]{0.49\linewidth} 
        \centering
        \includegraphics[width=\linewidth]{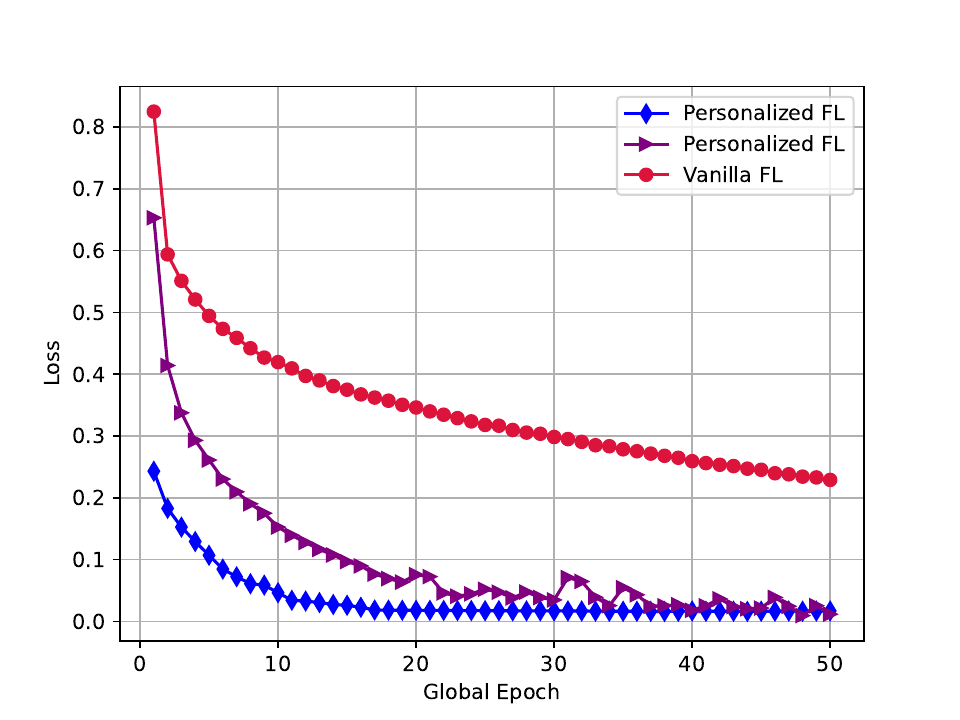}
        \caption{Average loss versus global communication rounds of FL model training for Cifar-10 dataset (non-IID).}
        \label{fig:sub4}
    \end{subfigure}
    \caption{Comparison of training convergence of our proposed PAFL scheme with personalized FL and existing multihop-FL schemes \cite{1,2}.}
    \label{fig:three_subs}
\end{figure}

Fig.~\ref{fig:three_subs} compares the convergence performance of our personalized learning and adaptive aggregation-aware FL (PAFL) model training with that of personalized FL and vanilla FL. Notably, this figure highlights the contrast between our proposed PAFL scheme and personalized FL method. Moreover, it also compares our PAFL scheme with the approaches in \cite{1,2}, which focus on vanilla FL. In Fig.~\ref{fig:sub1} and Fig.~\ref{fig:sub2}, we evaluate accuracy and loss across a series of global epochs for non-IID MNIST dataset, respectively. In Fig.~\ref{fig:sub1}, the PAFL approach consistently achieves higher accuracy, demonstrating its superior ability to adapt to the heterogeneous data distributions commonly encountered in real-world FL scenarios. This adaptability underscores its effectiveness in addressing the individualized needs of diverse nodes within the network. Fig.~\ref{fig:sub2} presents a similar trend, with PAFL showing more significant loss reduction compared to both personalized and vanilla FL. This further validates the enhanced performance of the PAFL approach. Fig.~\ref{fig:sub3} and Fig.~\ref{fig:sub4}, based on the non-IID CIFAR-10 dataset, exhibit trends consistent with those in Figs.~\ref{fig:sub1} and \ref{fig:sub2}. Specifically, PAFL maintains superior accuracy and loss reduction, confirming its effectiveness in scenarios with non-IID data distributions.

\begin{figure}[ht!]
    \centering
    \footnotesize
    \begin{subfigure}[t]{0.99\linewidth} 
        \centering
        \includegraphics[width=\linewidth]{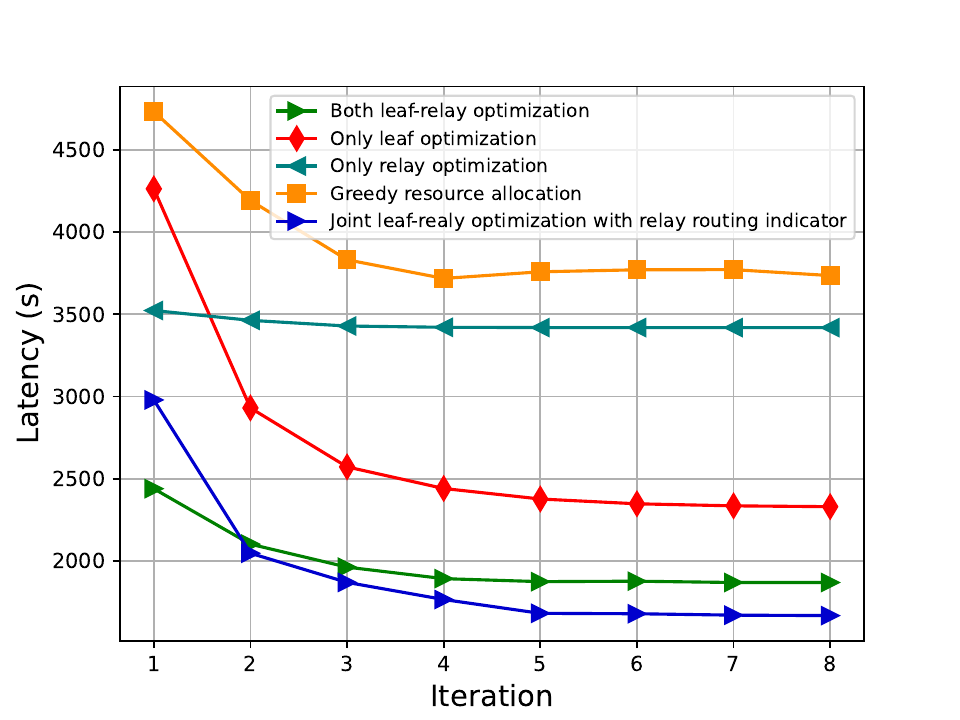}
        \label{fig:sub6}
    \end{subfigure}
    \caption{Latency comparison.}
    \label{fig:avloss}
\end{figure}

PAFL for MNIST dataset converges after 84 global rounds. Hence, we use $K=84$ in our latency optimization. Fig.~\ref{fig:avloss} depicts the latency (in seconds) versus the number of iterations, comparing our proposed algorithm with Scheme 1, Scheme 2, Scheme 3, and Greedy scheme. From the graph, it is evident that our devised scheme attains a consistent level of latency after the fifth iteration, significantly outperforming the other four schemes in terms of minimizing the latency level. Numerically, our proposed scheme achieves $19.79\%$, $45.33\%$, $13.16\%$, and $49.96\%$ lower latency compared to Scheme 1, Scheme 2, Scheme 3, and Greedy, respectively.


\begin{figure}[ht!]
    \centering
    \footnotesize
    \begin{subfigure}[t]{0.49\linewidth} 
        \centering
        \includegraphics[width=\linewidth]{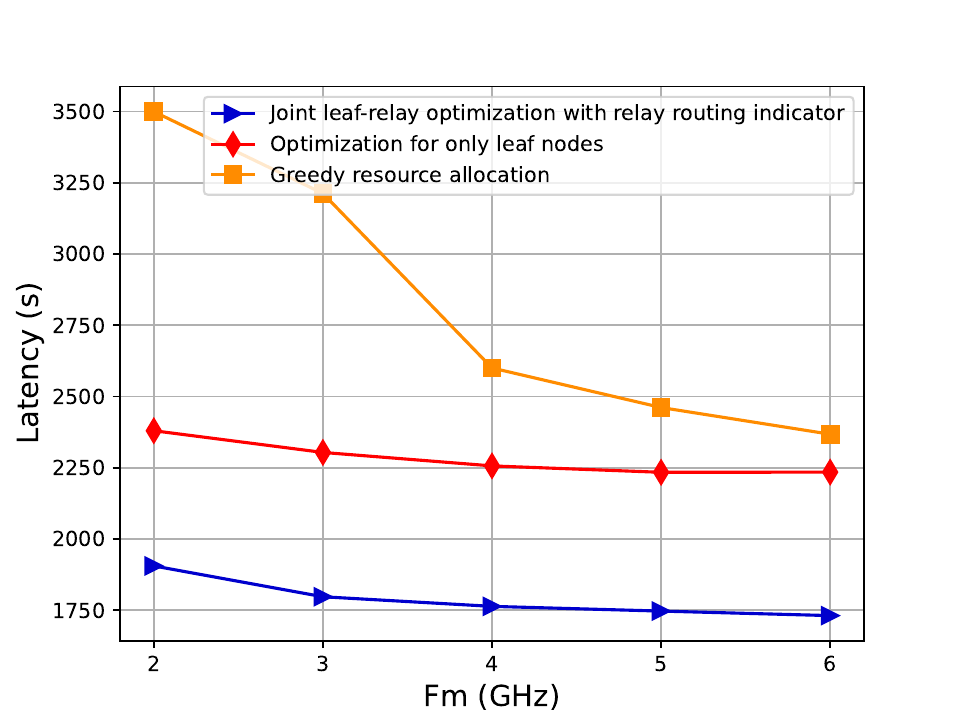}
        \caption{System latency versus maximum frequency of leaf nodes.}
        \label{fig:sub7}
    \end{subfigure}
    \hfill 
    \begin{subfigure}[t]{0.49\linewidth} 
        \centering
        \includegraphics[width=\linewidth]{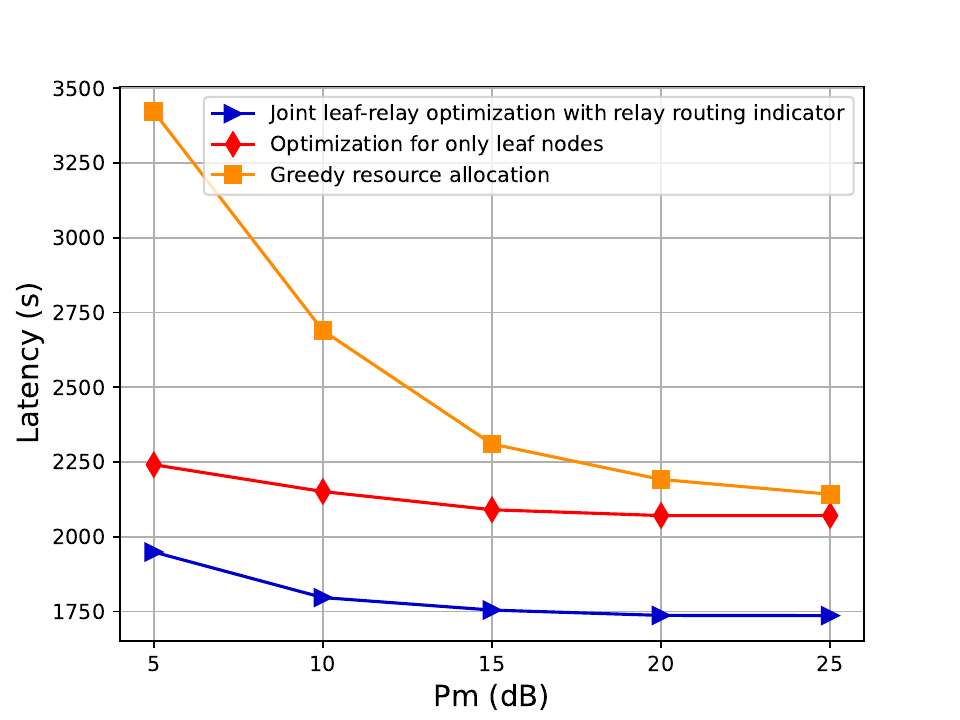}
        \caption{System latency versus maximum transmit power of leaf nodes.}
        \label{fig:sub8}
    \end{subfigure}
    \caption{Comparison of system latency with different schemes in terms of leaf nodes.}
    \label{fig:leaf}
\end{figure}
Moreover, we investigate the latency performance of different schemes.  Fig.~\ref{fig:sub7} illustrates the latency (in seconds) versus the maximum frequency (in GHz) of a leaf node where performance of our proposed algorithm has been compared with Scheme 1 and Greedy scheme. While all the schemes see a reduction in latency as the maximum frequency of leaf nodes increases, the proposed scheme exhibits approximately a notable $22.54\%$ and $36.78\%$ decrease in latency compared to Scheme 1 and Greedy scheme, respectively. Fig.~\ref{fig:sub8} shows the latency (in seconds) versus the maximum transmit power of a leaf node, comparing our proposed scheme with Scheme 1 and Greedy scheme. Our proposed scheme achieves approximately $16.15\%$ and $18.94\%$ lower latency compared to Scheme 1 and Greedy scheme, respectively, despite both schemes experiencing latency reduction with increased maximum transmit power of leaf nodes. Our proposed scheme demonstrates superior performance by dynamically adapting to network conditions through optimization of both leaf and relay nodes, along with the relay routing indicator. It allocates resources more effectively, outperforming both Scheme 1 and the traditional Greedy algorithm, which ignore broader network dynamics. Moreover, without relay routing optimization, relay nodes randomly share models with nearby nodes, which may not exist due to node departures. With optimization, our scheme reduces latency by enabling efficient routing, ensuring reliable and effective model relaying even in dynamic environments. 

\begin{figure}[ht!]
    \centering
    \footnotesize
    \begin{subfigure}[t]{0.49\linewidth} 
        \centering
        \includegraphics[width=\linewidth]{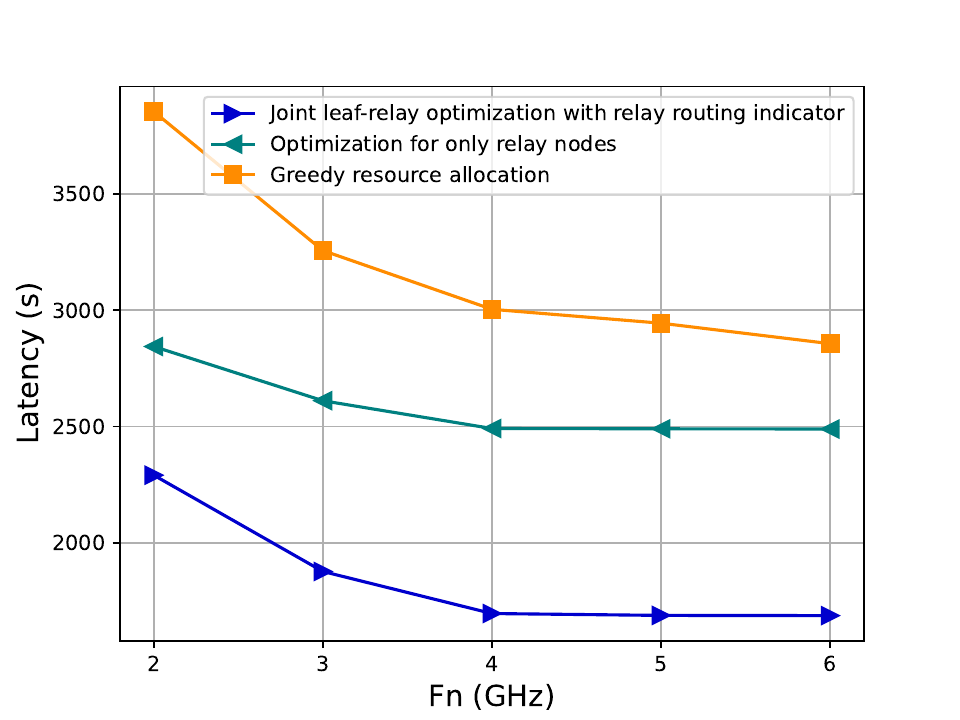}
        \caption{System latency versus maximum frequency of relay nodes.}
        \label{fig:sub9}
    \end{subfigure}
    \hfill 
    \begin{subfigure}[t]{0.49\linewidth} 
        \centering
        \includegraphics[width=\linewidth]{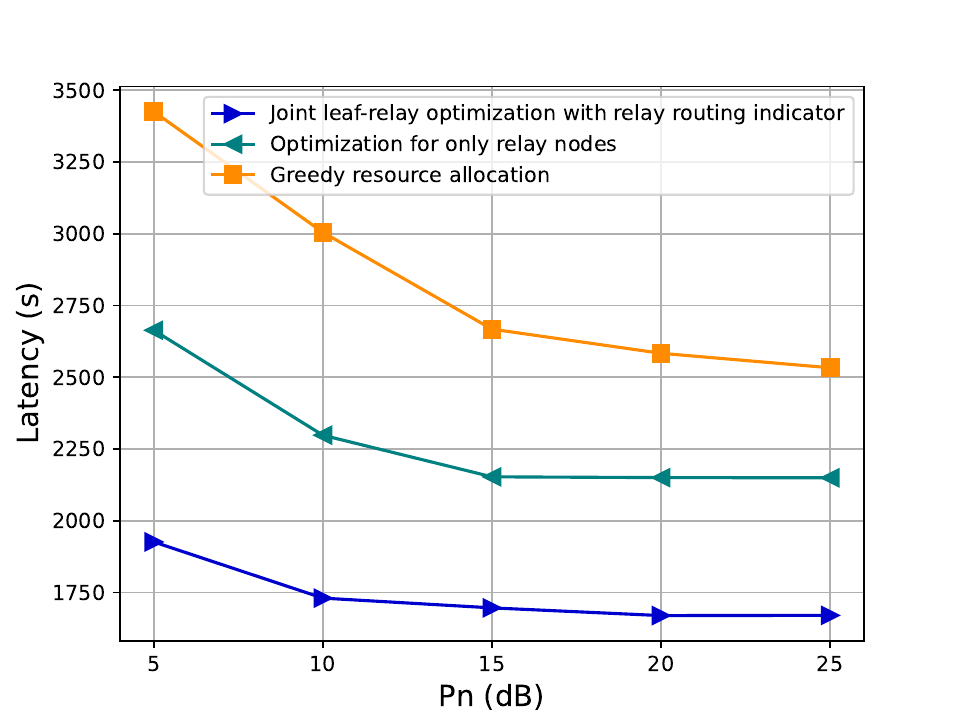}
        \caption{System latency versus maximum transmit power of relay nodes.}
        \label{fig:sub10}
    \end{subfigure}
    \caption{Comparison of system latency with different schemes in terms of relay nodes.}
    \label{fig:relay}
\end{figure}

Fig.~\ref{fig:sub9} illustrates the latency (in seconds) as a function of the maximum frequency (in GHz) of a relay node, comparing the performance of our proposed algorithm with Scheme 2 and the Greedy scheme. Similar to leaf node optimization, both Scheme 2 and the Greedy scheme exhibit a gradual reduction in latency as the maximum frequency of relay nodes increases. However, our proposed algorithm achieves significantly lower latency, reducing it by $32.25\%$ compared to Scheme 2 and by $69.37\%$ compared to the Greedy scheme. Fig.~\ref{fig:sub10} depicts the latency (in seconds) plotted against the maximum transmit power of a relay node,  further comparing the performance of Scheme 2, the Greedy scheme, and our proposed method. As with maximum frequency, increasing the maximum transmit power of relay nodes results in reduced latency for both Scheme 2 and the Greedy scheme. Nevertheless, our proposed algorithm consistently outperforms the alternatives, achieving $22.29\%$ lower latency than Scheme 2 and $51.64\%$ lower latency than the Greedy scheme. 

\begin{table}[ht]
\centering
\begin{tabular}{|c|c|c|}
\hline
\textbf{Number of Nodes} & \multicolumn{1}{c|}{\textbf{FL latency with EH}} & \multicolumn{1}{c|}{\textbf{FL latency without EH}} \\
\hline
3  & 2419.1160 & 2478.1008 \\
\hline
6  & 3051.3840 & 3165.5652 \\
\hline
9  & 3983.4564 & 4283.1684 \\
\hline
\end{tabular}
\caption{Comparison of latency with energy harvesting.}
\label{tab:latency_values}
\end{table}

In Table~\ref{tab:latency_values}, as the number of nodes increases, our proposed joint optimization method with the energy harvesting scheme consistently shows lower latency (in seconds), with the difference becoming more significant as the number of nodes grows. This indicates that energy harvesting provides substantial benefits in networks with more nodes.

\section{Conclusion}
In this paper, we minimized system latency for FL over multi-hop wireless  networks. The latency of the PAFL system was analyzed for both computation and communication delay. Frequency and transmit power of leaf and relay nodes have been jointly computed to minimize system latency via convex optimization, along with relay routing indicator. Through simulations, our approach can effectively reduce the latency of FL system (up to $69.37\%$ lower latency) in comparison to baselines.

\color{black}
\bibliographystyle{IEEEtran}
\bibliography{Main-Bibliography}

\end{document}